# Thin films with precisely engineered nanostructures

Rohit Sarkar[1] and Jagannathan Rajagopalan[1, 2*]

1. Materials Science and Engineering, School for Engineering of Matter Transport and Energy, Arizona State University, Tempe, AZ 85287, USA.
2. Mechanical and Aerospace Engineering, School for Engineering of Matter Transport and Energy, Arizona State University, Tempe, AZ 85287, USA.

Synthesis of rationally designed nanostructured materials with optimized mechanical properties, e.g., high strength with considerable ductility, requires rigorous control of diverse microstructural parameters including the mean size, size dispersion and spatial distribution of grains. However, currently available synthesis techniques can seldom satisfy these requirements. Here, we report a new methodology to synthesize thin films with unprecedented microstructural control via systematic, in-situ seeding of nanocrystals into amorphous precursor films. When the amorphous films are subsequently crystallized by thermal annealing, the nanocrystals serve as preferential grain nucleation sites and control their microstructure. We demonstrate the capability of this approach by precisely tailoring the size, geometry and spatial distribution of nanostructured grains in structural (TiAl) as well as functional (TiNi) thin films. The approach, which is applicable to a broad class of metallic alloys and ceramics, enables explicit microstructural control of thin film materials for a wide spectrum of applications.

[*] Corresponding author. Email: jrajago1@asu.edu

Controlling the fine-scale structure of materials is a fundamental objective of materials science as the microstructure predominantly influences a material's mechanical properties. With regard to structural materials, grain size refinement (up to the nanometer regime) has been used to significantly enhance the strength, fatigue resistance and tribological properties [1–3]. Unfortunately, the high strength and wear resistance of nanostructured materials generally come with substantial losses in toughness and ductility, which has limited their practical utility. Recent studies have shown that by altering the size dispersion and spatial distribution of grains at the nanoscale, a superior combination of mechanical properties, e.g., high strength with considerable ductility, can be achieved [4–6]. Yet, none of the currently available synthesis methods [7] provide the precise microstructural control required to realize rationally designed nanostructured materials with optimized properties.

Top-down methods like severe plastic deformation that are used to produce bulk nanostructured materials offer minimal capability to vary the grain size distribution or texture and often yield heterogeneous microstructures that exhibit large variation in mechanical properties [8]. Alternate bulk processing techniques like ball-milling and consolidation suffer from additional problems related to material purity and porosity [9]. Bottom-up methods provide a greater capability to control the microstructure compared to top-down methods. In electrodeposition, for instance, the mean grain size and twin density can be reproducibly controlled by altering current densities or using additives [10], even though impurity contamination is still an issue. Physical vapor deposition (PVD) processes, which are typically used to synthesize nanostructured thin films, are cleaner than electrodeposition and allow us to vary the texture and mean grain size by changing the film thickness, substrate material, deposition rate or temperature [11]. Nevertheless, the grain size of the films and their thickness cannot be easily decoupled and there is little control over the grain size dispersion or spatial distribution.

Hence, researchers have experimented with different methods to alter the various stages of microstructural evolution in thin films, which in the case of Volmer-Weber growth includes nucleation, island formation, coalescence and grain coarsening [12,13]. In particular, several studies have focused on controlling the grain nucleation process through the use of seed layers or isolated seed crystals. The seed layers/crystals have been found to alter the nucleation kinetics [14], enhance epitaxial growth [15], alter the grain structure [16] and aid the growth of highly oriented polycrystalline films [17]. However, these techniques have only been used to control a narrow set of microstructural parameters in specific materials and their broader applicability is unclear.

Here, we report a new methodology to synthesize nanostructured thin films with unprecedented microstructural control from amorphous precursor films. By systematic, in-situ deposition of seed layers, we first create a dispersion of crystalline nanoscale domains (seeds) in the amorphously grown films. We then exploit the ability of these seeds to preferentially nucleate grains during thermal annealing to rigorously control the crystallization process and the final microstructure of the films. We demonstrate the capability and generality of this approach by precisely tailoring the geometry, size dispersion and spatial distribution of nanostructured grains in structural (TiAl) as well as functional (TiNi) thin films. The approach is applicable to a broad class of amorphously grown metallic alloys and ceramics and enables the synthesis of thin films and coatings with tailored microstructures for a wide spectrum of applications including thermal barrier systems, cutting tools, biomedical implants and MEMS.

**Results**

The first step of the process (Fig. 1a) is the deposition of a relatively thick, amorphous layer ($\lambda$ = 25-500 nm) of a metallic alloy or ceramic at room temperature (RT). This can, for example, be accomplished by co-sputtering the constitutive elements of a metallic alloy. The next step is the deposition of a seed layer (0.5-2 nm thick) that has a crystalline structure at room temperature (RT). As this layer is very thin, it is non-contiguous and results in the formation of crystalline seeds. The size and average spacing ($\delta$) of these seeds can be controlled by varying the thickness of the seed layer or the deposition rate and temperature [11]. These steps are repeated until the desired overall film thickness is obtained. Finally, the amorphous matrix with the seeds is crystallized by annealing at high temperature to obtain the nanostructured film. Since both $\lambda$ and $\delta$ can be independently varied during each step of the growth process, the size, dispersion and spatial distribution of grains can be explicitly controlled.

We performed a series of systematic studies to test the viability of the process and uncover the effects of seed density and distribution on the microstructure of TiAl and TiNi thin films. The amorphous layers in the TiAl films had a nominal atomic composition of 45% Ti and 55% Al, whereas the TiNi films were equiatomic. The details of the film deposition process, annealing conditions and microstructural characterization are described in the methods section.

First, we verified that the thin seed layers indeed lead to the formation of well-dispersed, nanosized seeds (Fig. 1b). We confirmed the crystallinity of the seeds (Fig. 1c) using high resolution transmission electron microscopy (HR-TEM) and convergent beam electron diffraction (CBED), and verified their chemical composition using energy dispersive X-ray spectroscopy (Supplementary Fig. 1). It should be noted that the presence of well delineated seeds does not preclude the possibility of locally contiguous layers of the seed material. Similarly, only a fraction of the seeds, which are above a critical size and have appropriate crystal orientations to produce diffraction contrast, are visible in the bright-field TEM image. Nonetheless, the average size and spatial distribution of the seeds are significantly different for different seed layer thicknesses (Supplementary Fig. 2), which strongly suggests that seed formation is not controlled by defects on the surface of the amorphous layer.

The seeds were found to modify both the crystallization kinetics and the final microstructure of the films. When seeded and unseeded amorphous TiAl films were annealed at 550°C for a period of two hours, we observed partial crystallization in the seeded film, whereas the unseeded film remained completely amorphous (Supplementary Fig. 3). More interestingly, the seeded films exhibited a much smaller mean grain size ($d_m$) compared to the unseeded films when the annealing temperature was increased. Figures 2a and 2b show microstructures of TiAl films without and with Ti seeds annealed at 600°C for a period of four hours. The unseeded films had $d_m$ ~ 1 µm, whereas the seeded film had a $d_m$ = 42 nm. Notably, the seeded film retained its nanostructure ($d_m$ = 48 nm) even upon annealing at 750°C (Fig. 2c), which is approximately 0.6 times the melting temperature ($T_m$) of TiAl. This high thermal stability contrasts with the typical behavior of nanostructured metals, which show abnormal grain growth even at low temperatures [18,19]. The same microstructural stability was also observed when an Al seed layer was used instead (Supplementary Fig. 4).

It is noted that the Ti and Al seed layers change the overall composition of Ti and Al in the film by less than 1% and do not induce the formation of additional phases, which may alter the grain size. Both the unseeded and seeded films are primarily composed of the gamma phase, as revealed by X-ray diffraction (Supplementary Fig. 5). The same trends in $d_m$ were observed in the unseeded and seeded TiNi films as well but $d_m$ was significantly higher compared to TiAl films (Fig. 2d-f).

As indicated in Fig. 1, if the seeds act as preferential nucleation sites, the mean in-plane spacing between the grain nuclei should scale with $\delta$ and the number of grains across the thickness should scale with the number of seed layers. In effect, a larger $\delta$ and $\lambda$ should lead to a larger mean diameter and height, respectively, for the grains. To test this assumption, we systematically varied $\delta$ and $\lambda$ and analyzed the resultant microstructures of the films. To increase $\delta$ for a given seed layer thickness, we induced the seeds to coarsen and coalesce [11] by increasing the film temperature ($T_{seed}$) for a short period of time immediately after the seed layer was deposited. We note that $T_{seed}$ was sufficiently small (≤ 150°C) to avoid crystallization of the amorphous layer underneath. After this step, the film was allowed to cool back to RT and the subsequent amorphous layer was deposited. Figures 3a-c provide direct evidence that by altering $\delta$ the mean grain size can be systematically varied. The mean grain size of the TiAl films was least (42 nm) when no seed coalescence was induced and progressively increased with increasing $T_{seed}$, even though all other material parameters (type, number and thickness of seed layers, and film composition) and annealing conditions for crystallization were identical.

To ascertain the effect of λ, we synthesized films with different number of seed layers and analyzed their microstructures through cross-sectional TEM (Fig. 4). The TEM analysis revealed that when a single seed layer is used, the grains in the crystallized film are columnar and typically traverse the entire thickness (Fig. 4a, d). In contrast, when more than one seed layer was used, multiple grains with more equiaxed structure were observed across the film thickness (Fig. 4b, e). More importantly, the average number of grains across the thickness scaled with the number of seed layers, which provides further proof that the seeds act as preferential grain nucleation sites and control the microstructure.

Apart from the ability to explicitly tailor the grain diameters and geometry, the method also enables systematic control of the size dispersion and spatial distribution of grains. To demonstrate this capability, we varied λ and δ across the thickness to create films with specific gradient microstructures. For instance, the films in Fig. 4c and 4f are composed of smaller, equiaxed grains near the top and bottom and larger grains in the middle, whereas the film in Supplementary Fig. 6 has ultrafine grains ($d_m$ > 200 nm) in one half and nanocrystalline grains ($d_m$ < 50 nm) in the other half. As mentioned earlier, such gradient/multimodal microstructures enable high strength, ductility and fatigue resistance [4–6,20] and promote significant deformability even in materials that exhibit limited plasticity [21]. The highly heterogeneous microstructure will also considerably enhance the capability of nanostructured materials to recover inelastic strain and dissipate energy [22].

To investigate how the seeds alter the crystallization process, we carried out in-situ TEM annealing of TiAl films with and without seed layers. The experiments reveal that the seeds indeed act as preferential grain nucleation sites and a large number of grains are simultaneously nucleated in the seeded films as the temperature is increased (Supplementary Fig. 7a). Because the spacing between the grains is inversely related to their density, the average size of the grains when they impinge on each other is small. Furthermore, because the nucleated grains have a strong (111) texture (Supplementary Fig. 3), abnormal grain growth induced by surface energy minimization is restricted [23]. Therefore, grain growth after impingement is minimal. In contrast, the crystallization process was significantly different in the unseeded films, where only a small number of grains nucleated upon annealing. Typically, the spacing between the nuclei was in the order of micrometers, which led to a significantly larger mean grain size (Supplementary Fig. 7c, d). In other words, the crystallization of the unseeded metallic glass films was growth controlled and not nucleation controlled. Such growth controlled crystallization has been recently observed in other metallic glass thin films as well [24].

The results described above demonstrate the capability of the method to tune the mean size, geometry, size dispersion and spatial distribution of grains in thin films. TiAl and TiNi have been used to demonstrate viability in this study, but this approach can be extended to a broad class of metallic alloys and ceramics that can be deposited in the amorphous state. Examples of such materials include binary (CuZr, FeAl, NiAl, ZrTi), ternary (ZrCuAl, PdCuSi) and quaternary (ZrAlCuNi) alloys and ceramics such as SiC and WC [25–29]. In this context, it should be mentioned that materials synthesized by vapor deposition are farther from equilibrium compared to those obtained by liquid-to-solid processes, which are typically used to produce bulk metallic glasses. Thus, a larger spectrum of metallic alloys (over wider composition ranges) can be deposited as amorphous thin films [29]. It is also relevant to note that seed layers have been shown to influence the growth of ceramic nanostructures [30] and nanostructuring leads to enhanced hardness and fracture toughness in SiC and WC [31].

While we have focused on the control of grain size, structure and spatial distribution, the approach outlined here provides a rich design template to manipulate the microstructure of ceramic and metallic alloy films. For example, non-constituent elements can be used as seed layers to nucleate and stabilize metastable phases [32] that cannot be obtained by conventional processing routes. Note that many metals form isolated islands during the initial stages of film growth on amorphous substrates [33] and hence can be employed as seed layers. Likewise, seed layers can be used to precisely alter the local composition and nucleate additional phases that have complementary properties to the primary phase. For instance, by selectively using thicker Ti seed layers near the surface, the lamellar phase (higher Ti composition, better fracture toughness) can be nucleated to mitigate surface crack initiation in gamma TiAl films that have high specific strength and thermal conductivity [34]. Thus, this approach can enable the synthesis of novel thin film materials with unique microstructural features and physical properties, which are unattainable using current processing techniques.

Finally, in addition to sputtering, the approach outlined here is directly applicable to other PVD processes such as evaporation and, in principle, can be adapted to electrodeposition, provided amorphous microstructures can be obtained [35,36]. We anticipate that the relative simplicity of the approach and its broad applicability to different materials and deposition processes will considerably enhance its technological impact.

## Methods

Plasma enhanced chemical vapor deposition (PECVD) was used to deposit 60 nm thick $SiN_x$ onto 200 μm thick (100) silicon wafers. The $SiN_x$ was deposited to prevent inter-diffusion between the silicon substrate and the thin film during annealing. TiAl and TiNi thin films were then co-deposited on the $Si/SiN_x$ substrate by DC magnetron sputtering using an ATC Orion 5 sputter system manufactured by AJA International. These films were deposited using 99.999 % pure Ti, Al and Ni targets manufactured by Kurt J. Lesker company and were amorphous in the as-deposited state. The composition of the films was controlled by tuning the power of the individual sputtering guns. Rutherford backscattering spectrometry was used to determine the atomic composition of the films and X-ray diffraction was used to confirm the amorphous nature of the films.

The co-deposition of the films was carried out at room temperature at a base pressure of $5\times10^{-8}$ Torr and an argon deposition pressure of 3 mTorr. The co-deposition was stopped at specific periods and thin (0.5-2 nm thick) layers of pure Ti or Al was deposited on the amorphous TiAl and TiNi films. These thin Al or Ti layers were crystalline in nature and formed nanoscale domains (seeds), which acted as preferential grain nucleation sites when the amorphous films were crystallized by thermal annealing. After the deposition of the crystalline seed layer, the co-deposition of TiAl and TiNi was resumed so that the crystalline seeds were encapsulated by two amorphous layers. This process was repeated to obtain a film comprised of single or multiple seed layers and the desired thickness. The thickness of the seed layers and their number and spacing were systematically altered to study the effect of seeds on the final microstructure. After the deposition was complete, the films were either annealed under vacuum (< $10^{-6}$ Torr) in the sputtering chamber or annealed in a FEI Tecnai F20 transmission electron microscope (TEM) using a Gatan TEM heating holder. For annealing experiments inside the sputtering chamber, the temperature was increased to the final value within five minutes and held steady for the required amount of time. For the in-situ TEM annealing, the temperature was increased gradually to counteract the thermal drift of the specimen.

TEM specimens were fabricated out of $Si/SiN_x$ wafers having the TiAl/TiNi films on the top surface using microfabrication techniques. First, photoresist was spin coated on back of the wafer. An EVG620 mask aligner was then used to align the coated wafer with a glass mask containing the TEM specimen patterns and exposed to UV light. The exposed photoresist was then developed to complete the pattern transfer. An STS deep reactive ion etcher was then used to back etch the wafer to reveal the freestanding thin film TEM specimen. The plan-view microstructures of the films were then analyzed using a JEOL 2010F TEM while in-situ heating experiments were carried out using the FEI Tecnai F20 TEM. A FEI NOVA200 dual beam focused ion beam and scanning electron microscope was used to fabricate cross-sectional TEM samples using conventional lift out techniques. The mean grain size of the films was typically determined using the Scherrer formula from XRD peak profiles obtained using a PANalytical X'Pert PRO X-ray diffractometer. However, when the mean grain size was significantly larger than the film thickness, it was directly calculated from TEM bright-field images. It has been shown that the grain sizes calculated using XRD and TEM are comparable [37].

## Acknowledgements

This project was funded by the National Science Foundation (NSF) grants CMMI 1400505, DMR 1454109 and CMMI 1563027. The authors would like to gratefully acknowledge the use of facilities at the John M. Cowley Centre for High Resolution Electron Microscopy and at the Centre for Solid State Electronics Research at Arizona State University.

## Author Contributions

J.R. conceived the idea. J.R. and R.S. selected the materials for thin film synthesis and designed experiments. R.S. fabricated the films, performed annealing studies and characterized their microstructure. Both authors analyzed the data, discussed results and wrote the manuscript.

# Figures

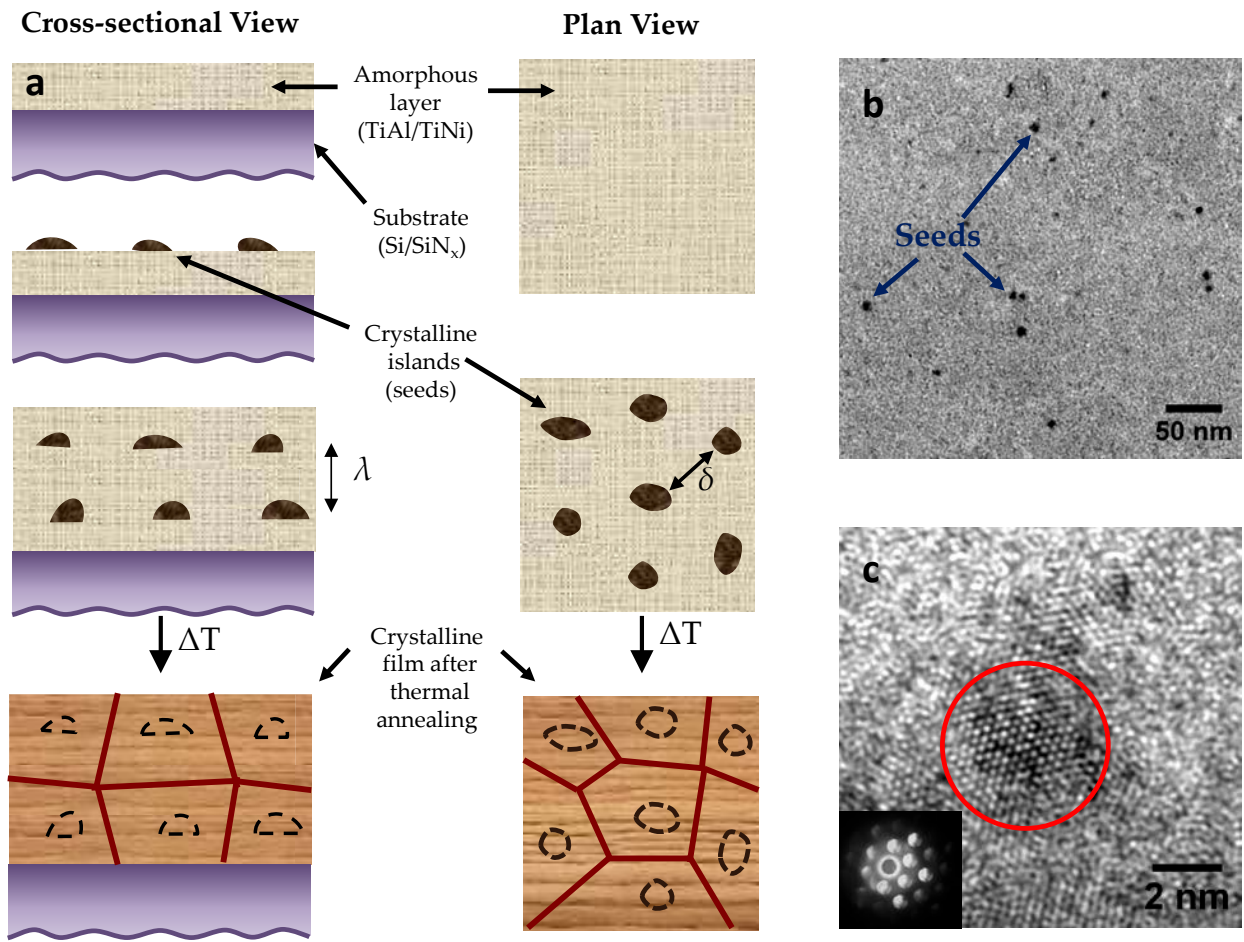

**Figure 1 Schematic of the process:** (a) First, an amorphous layer of the thin film material (TiAl/TiNi) is deposited on a substrate (Si/SiN$_x$). Next, a thin crystalline seed layer (Ti/Al) is deposited on the amorphous layer. The seed material is chosen such that this layer is non-contiguous and leads to the formation of crystalline nanoscale domains (seeds). A second amorphous layer is then deposited to sandwich the seeds. These steps are repeated until the desired film thickness is obtained. $\lambda$ is the spacing of the seed layers along the film thickness, while $\delta$ is the in-plane spacing between the seeds. Both $\lambda$ and $\delta$ can be independently varied during each step by altering the deposition parameters. Following the deposition, the film is thermally annealed to crystallize it. The seeds serve as grain preferential nucleation sites during annealing and hence the grain heights and diameters scale with $\lambda$ and $\delta$, respectively. (b) TEM bright-field image showing a dispersion of crystalline Ti seeds on an amorphous TiAl layer. (c) HR-TEM image showing the atoms on the (0001) plane of a Ti seed (red circle) surrounded by the amorphous TiAl matrix. The CBED pattern of the seed is shown in the inset.

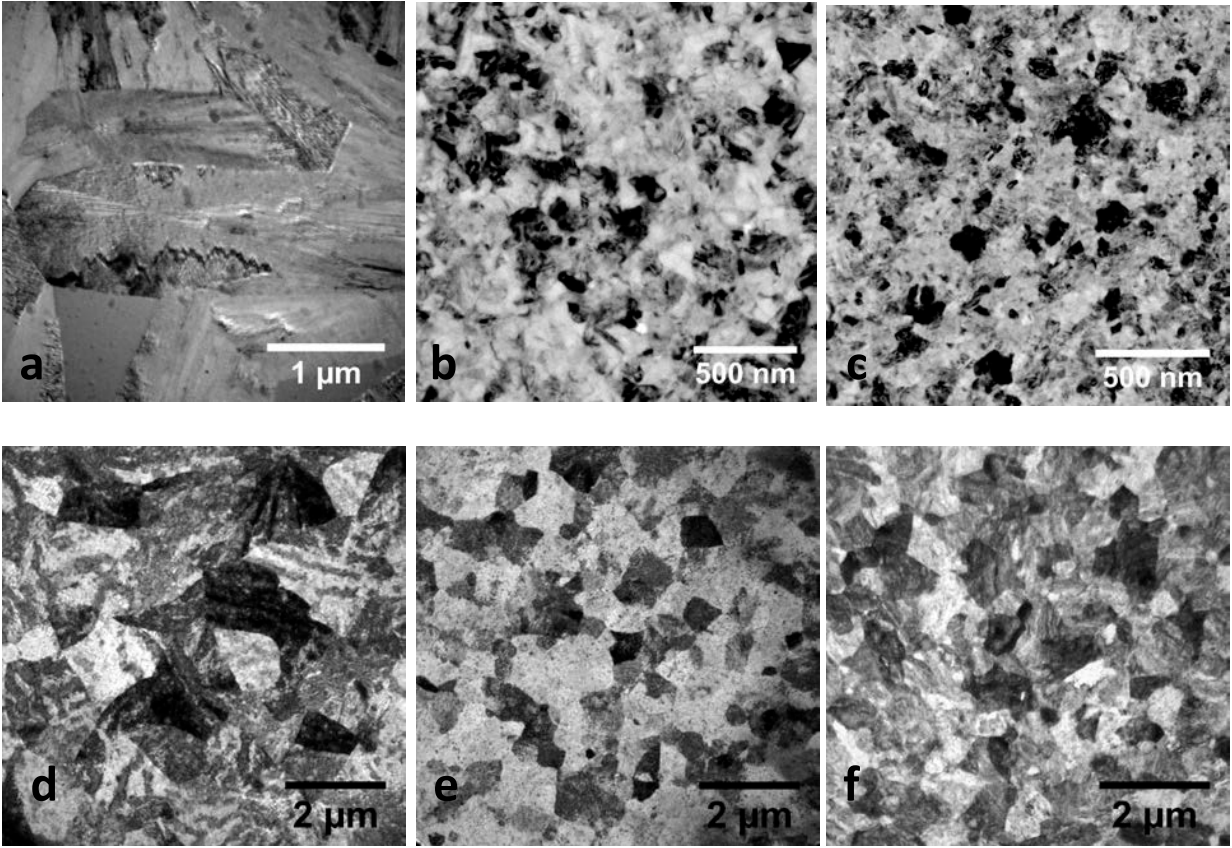

**Figure 2 Grain growth inhibition by seed crystals:** Bright-field TEM images of: (a) 100 nm thick TiAl film with no seed layer after annealing at 600°C for 4 hours showing large micrometer-sized grains ($d_m \sim 1$ μm). (b) 100 nm thick TiAl film with a single 1 nm Ti seed layer after annealing at 600°C for 4 hours ($d_m = 42$ nm). (c) 100 nm thick TiAl film with a single 1 nm Ti seed layer after 4 hours of annealing at 750°C. The grains remain in the nanocrystalline regime ($d_m = 48$ nm) and show little growth compared to (b). (d) 100 nm thick TiNi film with no seed layer after annealing at 600°C for 4 hours. Large micrometer-sized grains are visible and $d_m = 1.4$ μm. (e) 100 nm thick TiNi film with a single 1 nm thick Ti seed layer after annealing at 600°C for 4 hours ($d_m = 540$ nm). (f) 100 nm thick TiNi film with a single 1 nm thick Ti seed layer after annealing at 750°C for 4 hours. The film had a smaller mean grain size ($d_m = 580$ nm) compared to the unseeded film in (d) even at this temperature. For all the seeded films, the seed layer was deposited in the middle of the film.

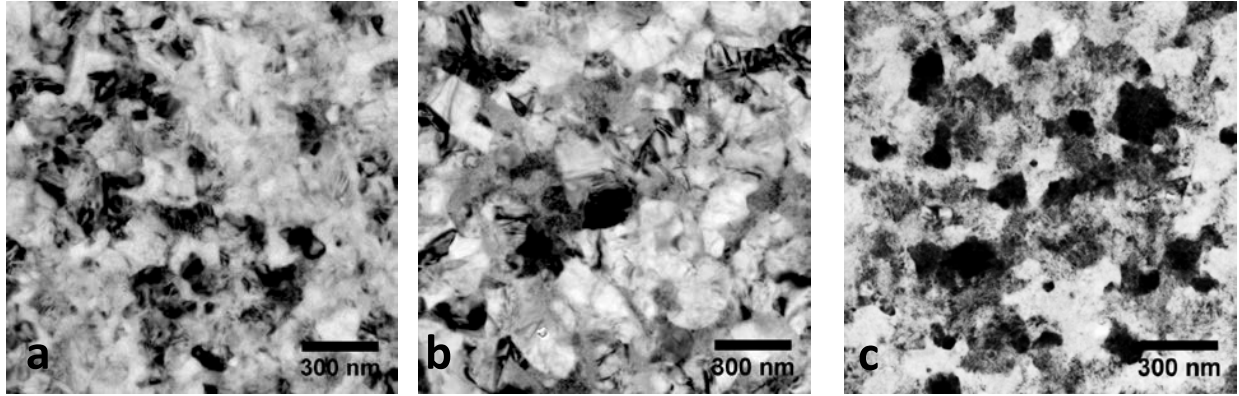

**Figure 3 Tailoring the mean grain size ($d_m$) of films:** (a-c) TEM bright-field images of three different 100 nm thick TiAl films with a single 1 nm Ti seed layer after annealing at 600°C for 4 hours. In addition to the final 600°C annealing, the films in (b) and (c) were subjected to a low temperature thermal anneal ($T_{seed}$) for 10 min immediately after the seed layer was deposited to increase the inter-seed spacing ($\delta$) by promoting seed coalescence. For the film in (a) which was not subjected to the low temperature anneal $d_m$ = 42 nm. The film in (b) was subjected to $T_{seed}$ = 100°C and $d_m$ increased to 95 nm. When $T_{seed}$ was increased to 150°C for the film in (c), $d_m$ further increased to 130 nm. In all the films, the seed layer was deposited in the middle.

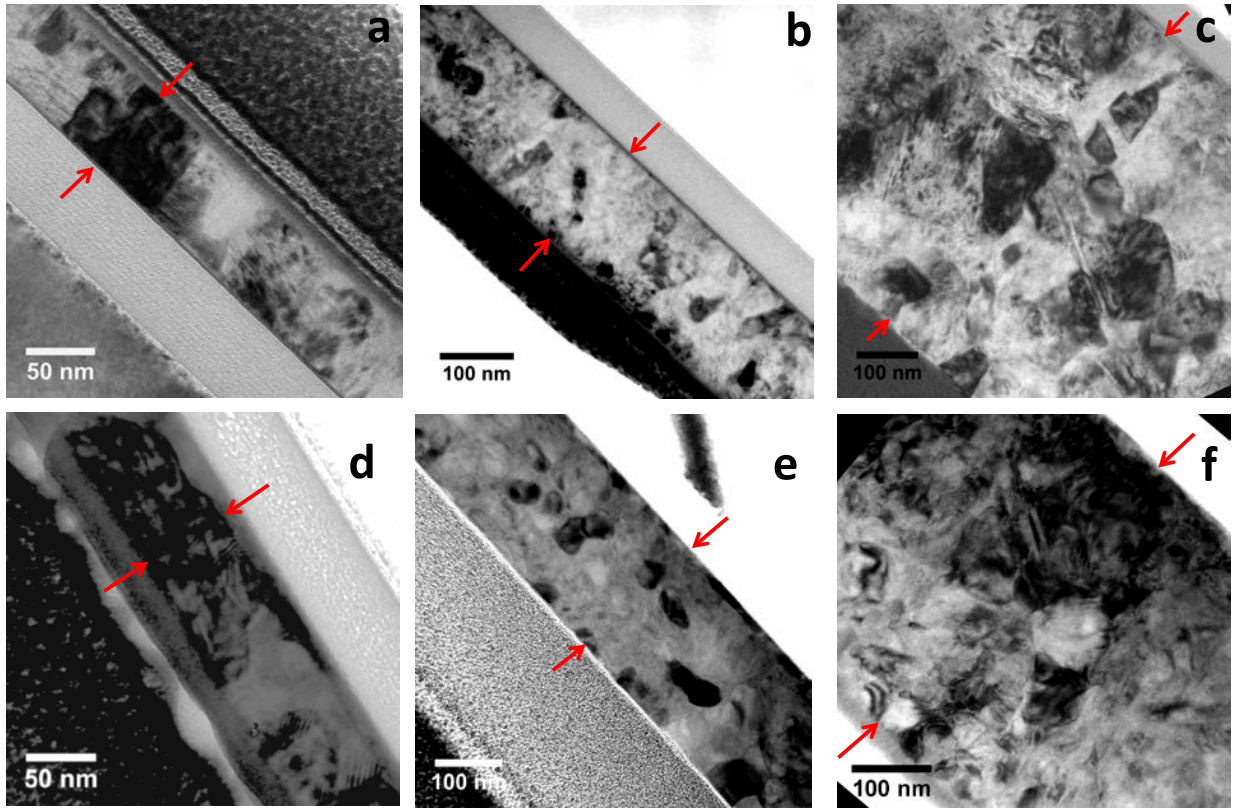

**Figure 4 Control of grain geometry, size dispersion and spatial distribution:** TEM bright-field cross-sectional images of: (a) A 100 nm thick TiAl film with a single 1 nm thick seed layer in the middle after annealing at 650°C for 4 hours. The grains typically traverse the entire thickness of the film and have a columnar structure. (b) A 200 nm thick TiAl film, with four 1 nm thick Ti seed layers spaced 50 nm apart, after annealing at 650°C for 4 hours. Multiple grains along the thickness are observed. (c) A TiAl film, with smaller seed layer spacing near the top and bottom and larger spacing in the middle, after annealing at 650°C for 4 hours. A gradient microstructure is formed upon annealing. (d) A 100 nm thick TiNi film with a single 1 nm thick Ti seed layer in the middle after annealing at 650°C for 4 hours. Similar to (a) single grains traverse the thickness of the film. (e) A 200 nm thick TiNi film, with four 1 nm thick Ti seed layers, after annealing at 650°C for 4 hours. Multiple grains along the thickness are observed. (f) A TiNi film, with multiple seed layers and varying $\lambda$, after annealing at 650°C for 4 hours. A gradient in the grain size can be seen along the thickness similar to (c). The red arrows in the figures indicate the TiAl and TiNi film cross-sections.

*Supplementary Information*

# Thin films with precisely engineered nanostructures

Rohit Sarkar[1] and Jagannathan Rajagopalan[1, 2]

1. Materials Science and Engineering, School for Engineering of Matter Transport and Energy, Arizona State University, Tempe, AZ 85287, USA.
2. Mechanical and Aerospace Engineering, School for Engineering of Matter Transport and Energy, Arizona State University, Tempe, AZ 85287, USA.

**Supplementary Figures**

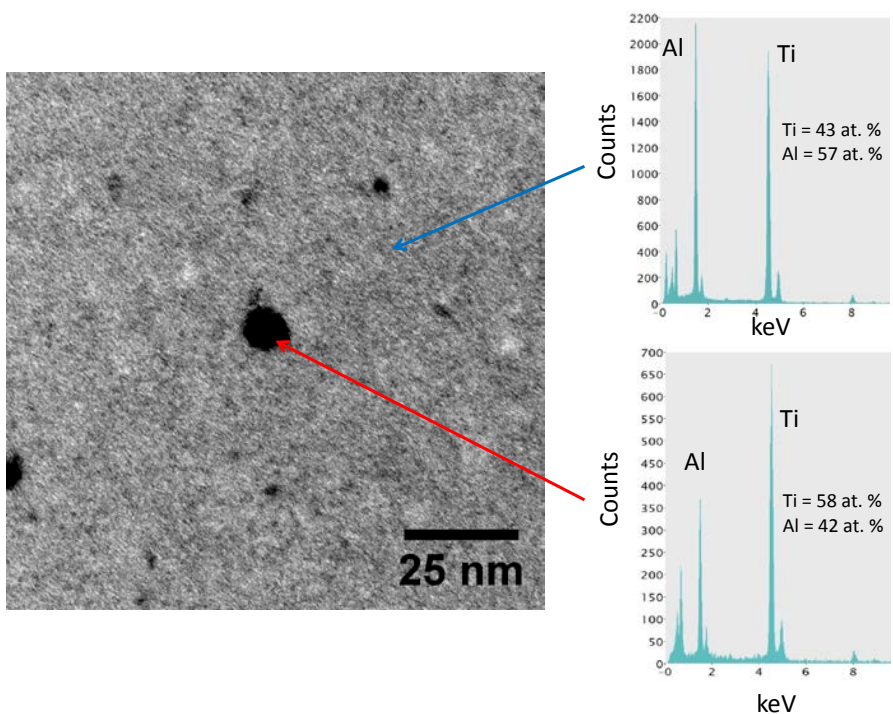

**Figure 1: Composition of seed crystals:** TEM bright-field image of a 40 nm thick amorphous TiAl film with a 1 nm thick Ti seed layer in the middle, before annealing. Isolated Ti seeds are clearly visible. Energy dispersive X-ray spectroscopy (right) of the seed region (indicated by red arrow) and the amorphous matrix (indicated by blue arrow) reveals the difference in their composition. Note that the composition of the seed region also includes the amorphous matrix encapsulating the seed. The accuracy of the composition determined by the method is ±2 %.

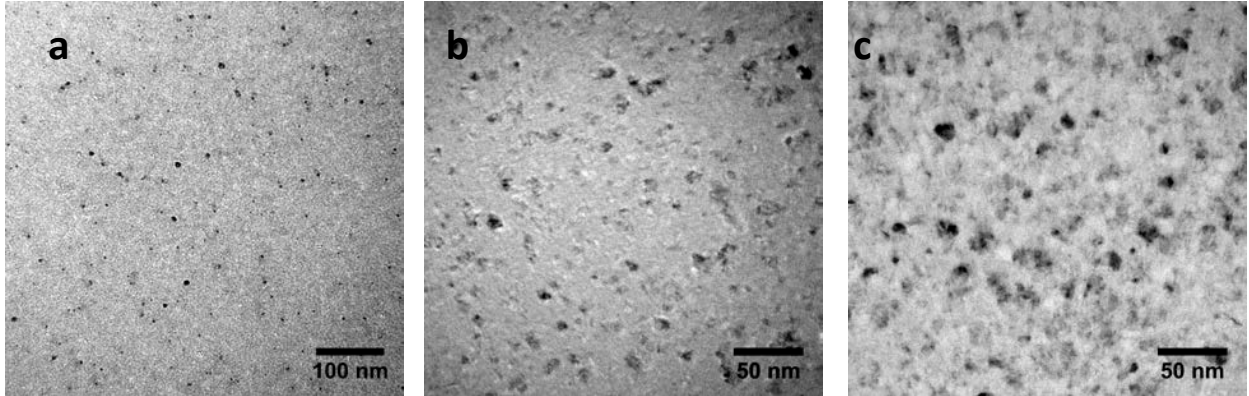

**Figure 2: Effect of seed layer thickness on size and spatial distribution of seeds:** TEM bright-field images of (a) 40 nm thick TiAl film with 0.5 nm thick Ti seed layer in the middle before annealing. (b) 40 nm thick TiAl film with 1.5 nm thick Ti seed layer in the middle before annealing. (c) 40 nm thick TiAl film with a 2 nm thick seed layer in the middle before annealing.

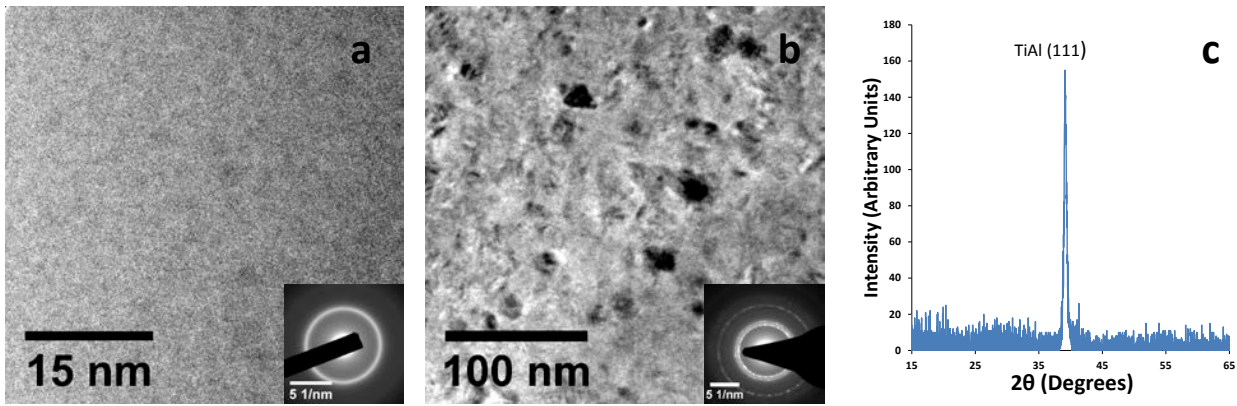

**Figure 3: Enhanced crystallization due to seeds:** (a) TEM bright-field image of a TiAl film with no seed layer after 2 hours of annealing at 550°C. This film remains completely amorphous as indicated by the diffuse ring in the selected area diffraction (SAD) pattern (inset). (b) TEM bright-field image of the TiAl film with a single 1 nm thick Ti seed layer in the middle after 2 hours of annealing at 550°C. The film showed partial crystallization and spots started to appear in the SAD pattern (inset). (c) XRD pattern of the film in (b). The presence of a well-defined (111) ϒ-TiAl peak indicates that the film has started to crystallize.

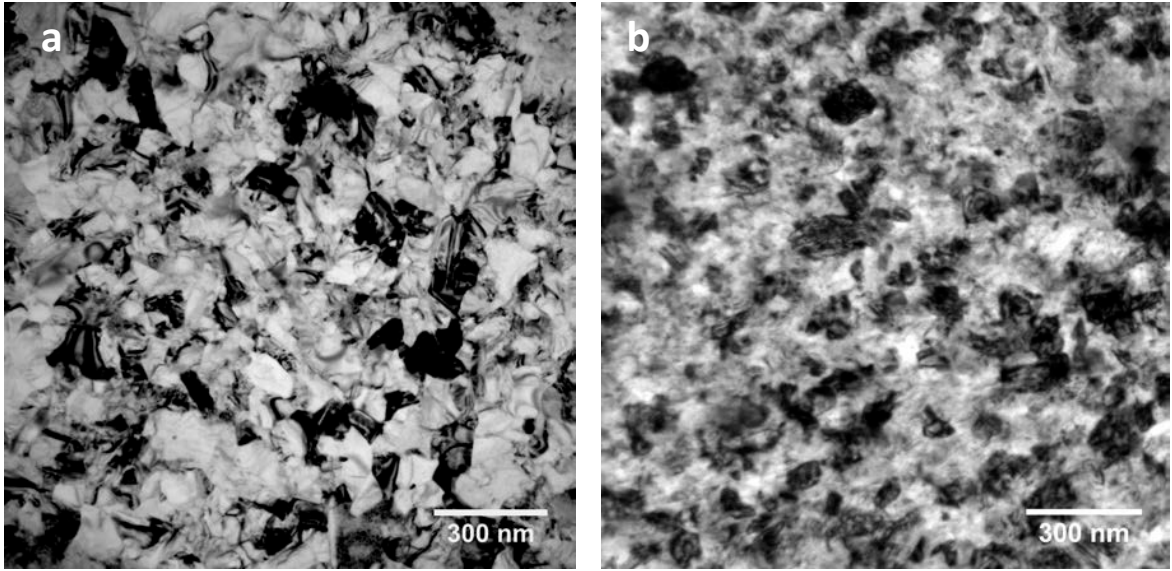

**Figure 4: TiAl films with Al seeds:** TEM bright-field image of (a) TiAl film, 100 nm thick with a single 1 nm Al seed layer in the middle after 4 hours of annealing at 650°C. The film was nanostructured at this temperature. (b) TiAl film, 100 nm thick with a single 1 nm Al seed layer in the middle after 4 hours of annealing at 750°C. The film retained its nanostructure even at this elevated temperature.

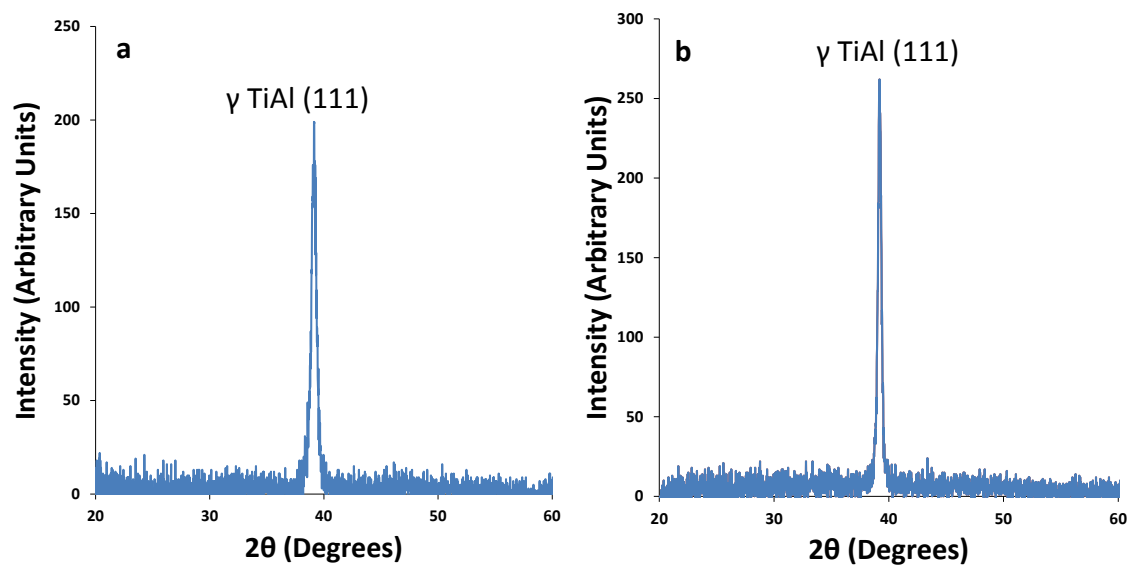

**Figure 5: Texture of TiAl films:** (a) XRD measurements show the strong (111) texture of a TiAl film with a 1 nm Al seed layer after 4 hours of annealing at 650°C. (b) XRD measurements of a TiAl film deposited without any seed layer after 4 hours of annealing at 600°C. This film also exhibits a strong (111) texture.

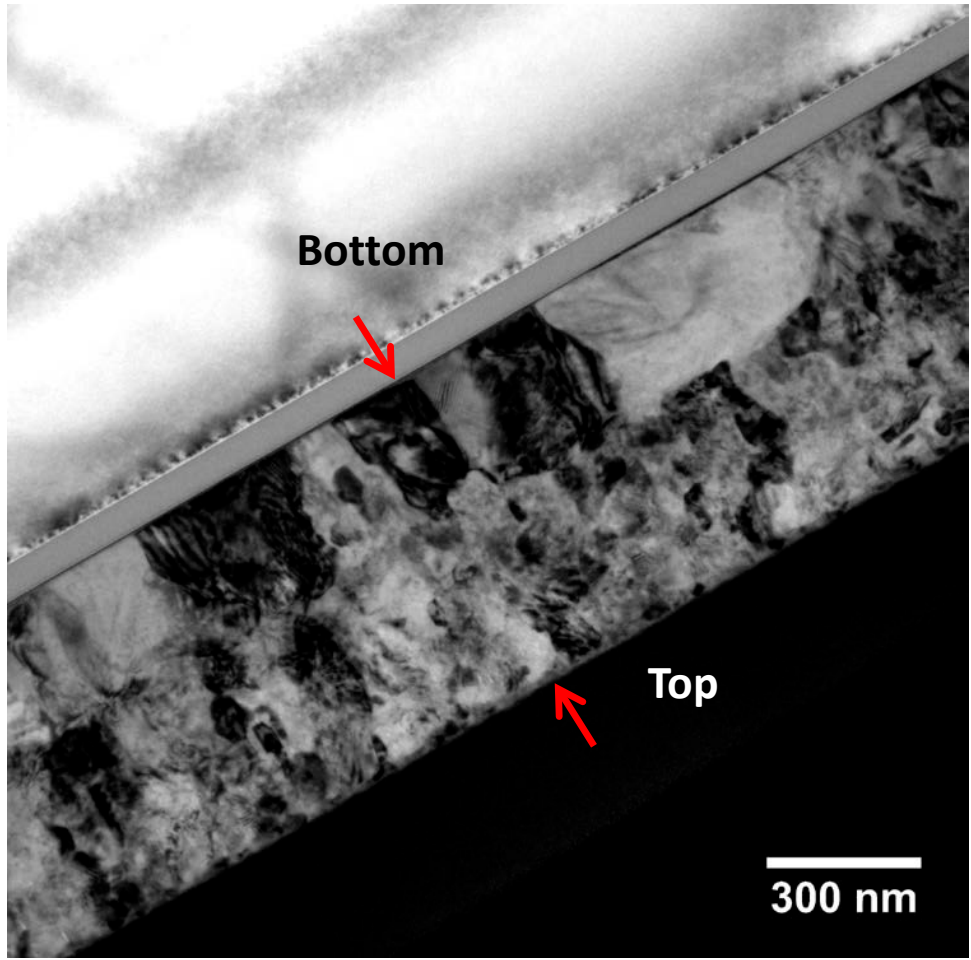

**Figure 6: TiAl film with gradient microstructure:** Cross-sectional TEM image of a TiAl film which was grown to have ultrafine grains (> 200 nm) in the bottom half and nanocrystalline grains (< 50 nm) in the top half. The gradient microstructure was obtained by having a much larger seed layer spacing ($\lambda$) in the bottom half of the film compared to the top half.

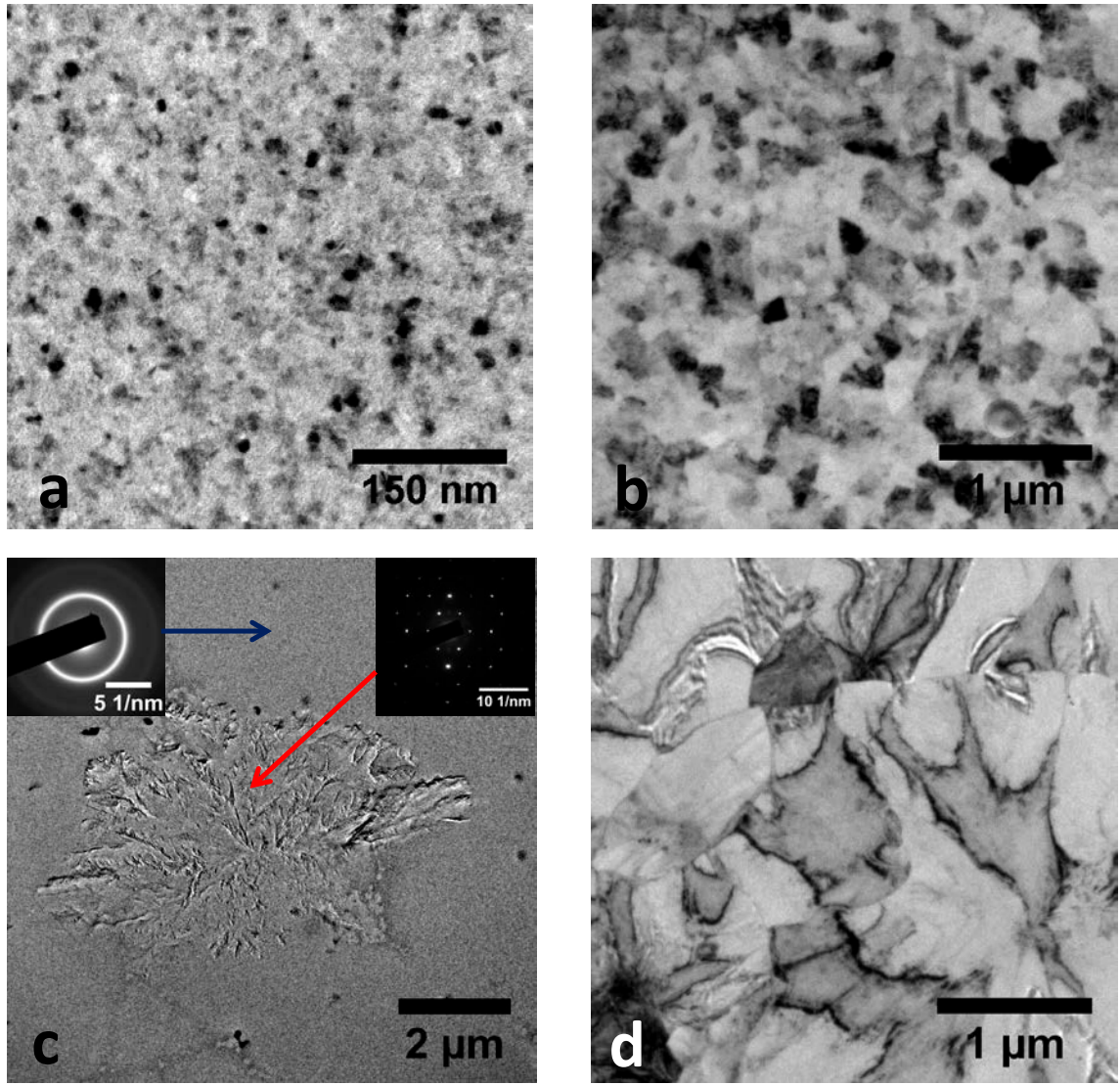

**Figure 7 In-situ TEM annealing:** (a-b) TEM bright-field images taken during in-situ annealing of a 100 nm thick TiAl film with a single 1 nm thick Ti seed layer in the middle. The image in (a) was taken after heating the film at 550°C for 1 hour and shows small grains with sizes ranging from 5-25 nm. Figure (b) corresponds to the film after heating at 650°C for 30 minutes and shows nanostructured grains. (c-d) TEM bright-field images taken during in-situ heating of a 100 nm thick TiAl film with no seed layer. The image in (c) was taken after heating the film at 600°C for 1 hour and shows the nucleation of a large crystalline phase (red arrow), as revealed by the corresponding selected area diffraction pattern. The rest of the film (blue arrow) still remained amorphous. Figure (d) corresponds to the film after heating at 650°C for 30 minutes and shows the formation of large micrometer-sized grains.